\documentstyle[psfig]{mn}

\def\phs{\phantom{1}}
\def\phd{\phantom{11}}
\def\phps{\phantom{.1}}
\def\php{\phantom{.}}
\def\phu{\phantom{$^1$}}

\title[`Double-Double' Radio Galaxies]{Radio galaxies with a
`double-double' morphology:\\
{\Large \bf I - Analysis of the radio properties and evidence for interrupted activity in active galactic nuclei}}
\author[A.P. Schoenmakers et al.]{Arno P. Schoenmakers$^{1,2,3}$\thanks{Current address: NFRA, P.O. Box 2,
    7990 AA, Dwingeloo, The Netherlands; Email: Schoenmakers@nfra.nl}, A.G. de
Bruyn$^{3,4}$, H.J.A. R\"{o}ttgering$^2$, H. van der Laan$^1$\newauthor 
and C.R. Kaiser$^{5,6}$\\ 
$^1$ Astronomical Institute, Utrecht University, P.O. Box 80\,000, 3508~TA Utrecht, The Netherlands\\
$^2$ Sterrewacht Leiden, Leiden University, P.O. Box 9513, 2300~RA Leiden, The 
Netherlands\\
$^3$ NFRA, P.O. Box 2, 7990~AA Dwingeloo, The Netherlands\\
$^4$ Kapteyn Astronomical Institute, University of Groningen, P.O. Box 800, 9700~AV Groningen, The Netherlands\\
$^5$ University of Oxford, Department of Physics, Nuclear Physics
Laboratory, Keble Road, Oxford, OX1 3RH, UK\\
$^6$ Max-Planck-Institut f\"{u}r Astrophysik, Karl-Schwarzschild-Str. 1,
85740 Garching bei M\"{u}nchen, Germany}
\date{Received ; accepted}
\pubyear{2000}

\begin{document}
\maketitle

\begin{abstract}
\noindent
We present four Mpc-sized radio galaxies
which consist of a pair of double-lobed radio sources, aligned along
the same axis, and with a coinciding radio core.  We have called these
peculiar radio sources `double-double' radio galaxies (DDRG) and
propose a general definition of such sources: A `double-double' radio
galaxy consists of a pair of double radio sources with a common
centre. Furthermore, the two lobes of the inner radio source must have
a clearly extended, edge-brightened radio morphology.  Adopting this
definition we find several other candidate DDRGs in the literature.
We find that in all sources the smaller (inner) pair of radio
lobes is less luminous than the larger (outer) pair, and that the
ratio of 1.4-GHz flux density of these two pairs appears to be
anti-correlated with the projected linear size of the inner source.
Also, the outer radio structures are large, exceeding 700 kpc.  We
discuss possible formation scenarios of the DDRGs, and we
conclude that an interruption of the jet-forming central activity 
is the most likely mechanism.  For one of our sources
(B\,1834+620) we have been able to observationally constrain the
length of time of the interruption to a few Myr. We discuss
several scenarios for the cause of the interruption and suggest
multiple encounters between interacting galaxies as a
possibility. Finally, we discuss whether such interruptions help the
formation of extremely large radio sources.

\end{abstract}

\begin{keywords}
galaxies: active -- galaxies: individual: B\,0925+420, B\,1240+389, B\,1450+333, B\,1834+620 -- galaxies: jets -- radio continuum: galaxies
\end{keywords}

\section{Introduction}
\label{sec:intro}
One of the outstanding issues concerning extragalactic radio sources
and other Active Galactic Nuclei (AGN) is the total duration of their
active phase. For radio sources, this physical age of the nuclear
activity is not to be confused with the radiative loss age determined
from radio spectral ageing arguments; many extragalactic radio sources
probably have a physical age well surpassing their radiative loss age
(e.g. van der Laan \& Perola 1969, Eilek 1996).  The length of the
active phase is intimately related to the possible existence of duty
cycles of nuclear activity. In case nuclear activity is not
continuous, how often do interruptions occur and how long do they
last?

AGN activity is believed to be associated with the presence of a
massive black hole (MBH) in the centre of a galaxy.  In the last few
years, observational evidence for the presence of such MBHs in nearby
galaxies has steadily increased (e.g. van der Marel
1998). Potentially, all these MBHs have the ability to invoke AGN
activity; the fact that most are not active (or `dormant') probably
results from a lack of fuel to drive such activity. An important
question is whether all galaxies harbouring a MBH went through one or
more periods of AGN activity.  Franceschini, Vercellone \& Fabian
(1998) find that multiple periods of activity are not ruled out on
basis of currently available data on MBHs. The time scale for such
recurrent activity would be very large, though, of order $10^9$ or
$10^{10}$ yr.

A duty cycle can only be recognized as such if there is some
mechanism to preserve the information of past nuclear activity for a
long enough time to be recognized when a new cycle starts up.  In
extended radio sources, such a mechanism is provided for by the radio
lobes, which are large reservoirs of energy resulting from very
powerful jet-producing AGN and which can potentially store information
of past activity for a long time after the central activity has
stopped.  If a new phase of activity should start before these `old'
radio lobes have faded, and if this activity manifests itself by the
production of jets, we can in principle recognize this by the
observation of a new, young radio source embedded in an old, relic
structure.

One well-known candidate for such a `restarted' radio source is the
radio galaxy 3C\,219 (Clarke \& Burns 1991, Clarke et al. 1992, Perley
et al. 1994). In this source radio jets have been observed which
abruptly become undetectable at some point between the core and the
leading edge of the outer radio lobes. Clarke et al. (1992) proposed
that the jets in this source could be restarting, but numerical
simulations of this mechanism predict structures that have not (yet)
been observed (e.g. Clarke \& Burns 1991). Bridle et al. (1989)
suggested that a restarting jet may also be the cause of some of the
peculiar properties of the radio galaxy 3C\,288. Since in these sources
the new jet activity must have started relatively short after the halting of
the old jet activity, we prefer to use the term `interrupted
activity' here, as opposed to the above-mentioned `recurrent activity'
which relates to activity phases separated by much larger time spans. 

This paper is the first in a series of three related to possible observed
interrupted activity in large radio sources. Paper II (Kaiser, Schoenmakers
\& R\"{o}ttgering 1999)
presents a detailed model for the evolution of radio sources with
restarted jets in which the direction of the jet flow between subsequent
periods of activity is similar. In Paper III (Schoenmakers et al. 1999b) 
we present the results of a detailed observational study of one the
sources introduced in this paper, B\,1834+620 (see also Lara et al. 1999).

Here, we discuss four peculiar radio galaxies with morphologies that
strongly suggest the occurence of interrupted radio-activity. In Section
2 we will present radio and optical data of these four sources. In
Section 3 we will argue why these sources form a separate class of
radio source, which we will designate as the the class of
`double-double' radio galaxies.  We will add a further three sources
from the literature which closely resemble the four sources introduced
in Section 2. Section 4 contains a brief analysis of their radio
properties. Section 5 then discusses possible causes for the observed
morphology. We will argue that a restarted jet is the most likely
scenario for the formation of the inner structure of these radio
sources. Further, we will discuss what may have caused the
interruption in these sources and what other indications there are for
interrupted AGN activity. Finally, we will discuss if the formation of
extremely large radio sources may be related to this phenomenon.
Our conclusions can be found in Section 6.

We adopt $H_0 = 50$\,km\,s$^{-1}$\,Mpc$^{-1}$ and $q_0=0.5$
throughout this paper. A spectral index $\alpha$ is defined according to
the relation $S_{\nu} \propto \nu^{\alpha}$, where $S_{\nu}$ is the flux
density at frequency $\nu$.

\begin{figure*}
\psfig{figure=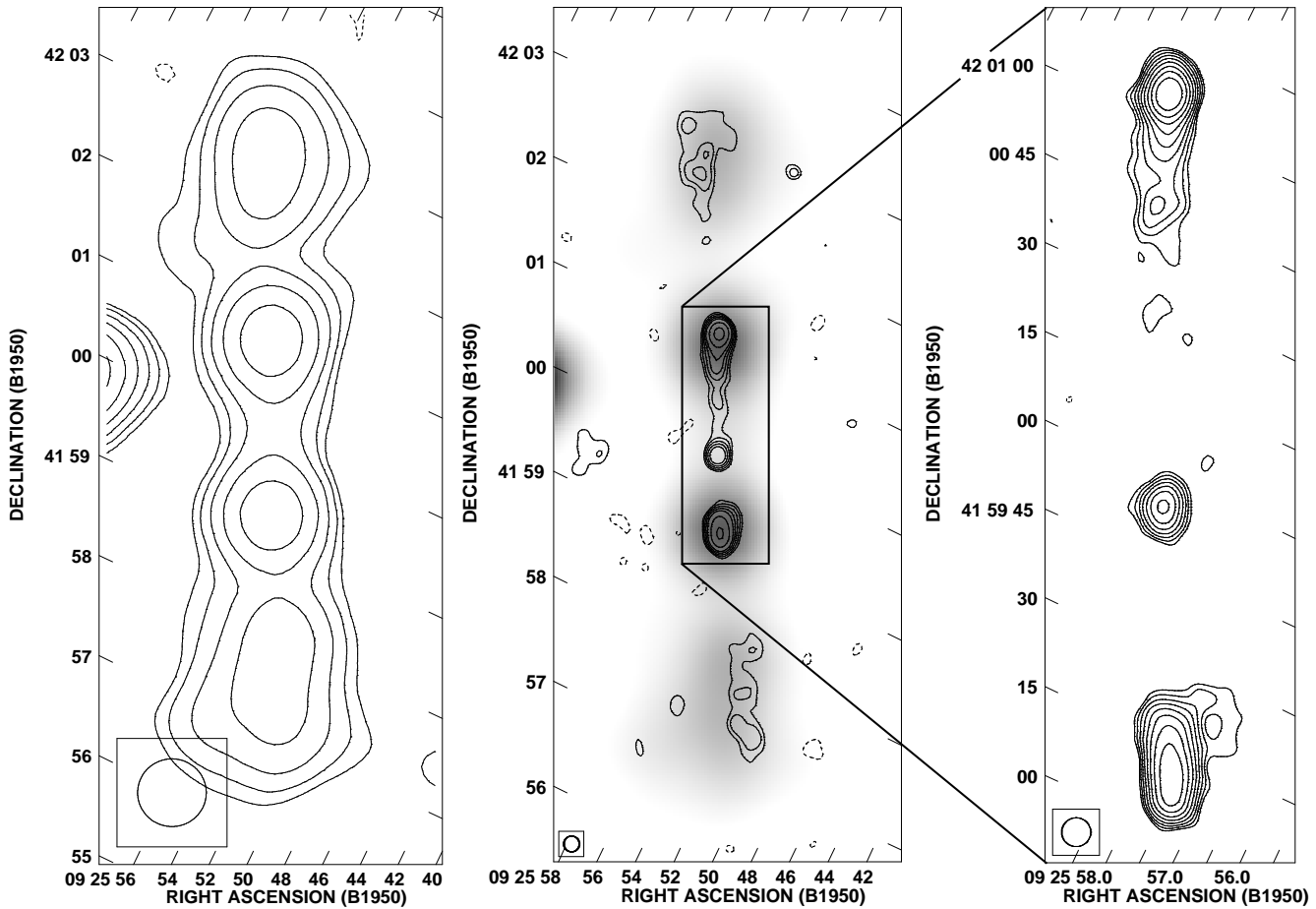,width=\textwidth,angle=270}
\caption{\label{fig:0925}Radio contour plots of the source
B\,0925+420, rotated clock-wise (CW) by $26\degr$. The FWHM size of
the restoring beam is indicated in the panel at the lower left side of
each plot. {\bf a} Plot from the 1.4-GHz NVSS survey. Contours are at
($-1.3$,1.3,2.6,5.2,10.4,20.8,41.6,93.2) mJy beam$^{-1}$. {\bf b} Plot
from the 1.4-GHz FIRST survey, convolved to a beamsize of $10\arcsec$
(FWHM).  Contours are at ($-0.8$,0.8,1.13,1.6,2.26,3.2,6.4,12.8) mJy
beam$^{-1}$. The greyscale represents the flux density distribution in
the NVSS radio map. {\bf c} Contourplot of the inner radio structure
from the FIRST survey at full resolution ($5\farcs4$ FWHM). Contours
are at ($-0.45$,0.45,0.64,0.9,1.27,1.8,2.55,3.6,5.09,7.2) mJy
beam$^{-1}$.}
\end{figure*}

\begin{figure*}
\psfig{figure=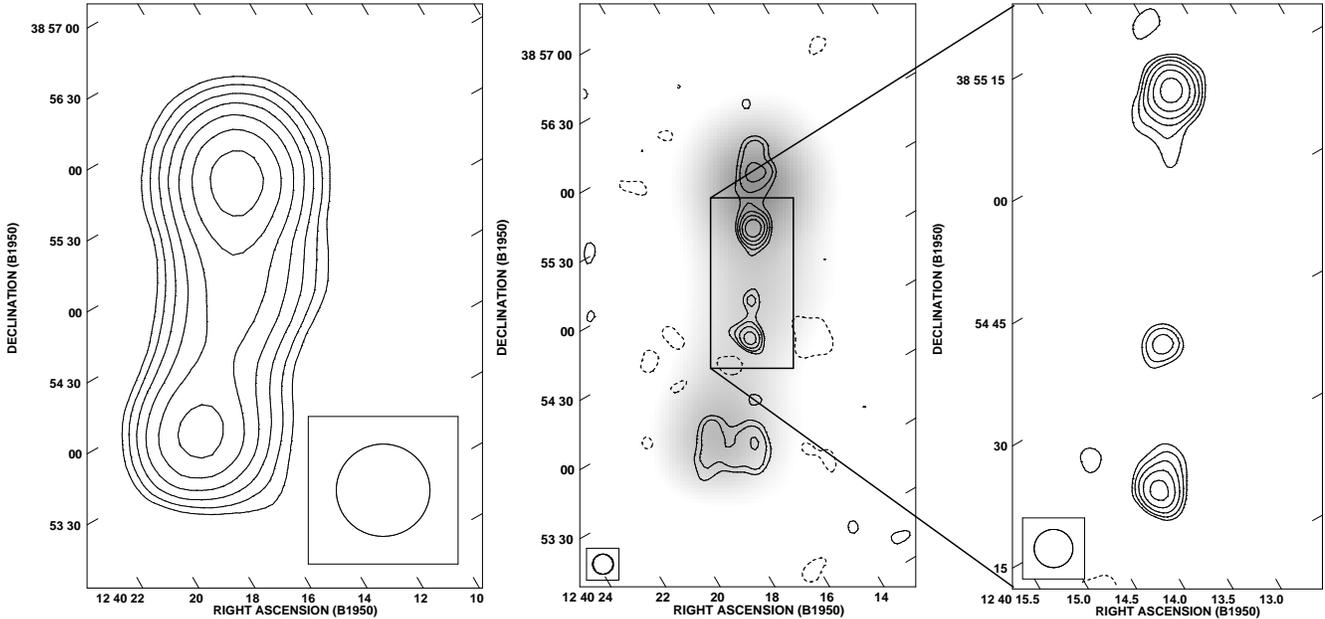,width=\textwidth,angle=270}
\caption{\label{fig:1240}Radio contourplots of the source
B\,1240+389. The plots have been rotated counter clock-wise (CCW) by
$23\degr$. The FWHM size of the restoring beam is indicated in the
panel at the lower left, or right, side of each plot. {\bf a} Plot
from the 1.4-GHz NVSS survey. Contours are at
($-1.2$,1.2,1.7,2.4,3.4,4.8,6.8,9.6) mJy beam$^{-1}$. {\bf b} Plot
from the 1.4-GHz FIRST survey convolved to a beamsize of $10\arcsec$
(FWHM).  Contours are at ($-0.6$,0.6,0.85,1.2,1.7,2.4) mJy
beam$^{-1}$. The greyscale represents the flux density distribution in
the NVSS radio map. {\bf c} Contourplot of the inner radio structure
from the FIRST survey at full resolution.  Contours are at
($-0.4$,0.4,0.56,0.8,1.13,1.6,2.26) mJy beam$^{-1}$.}
\end{figure*}

\begin{figure*}
\psfig{figure=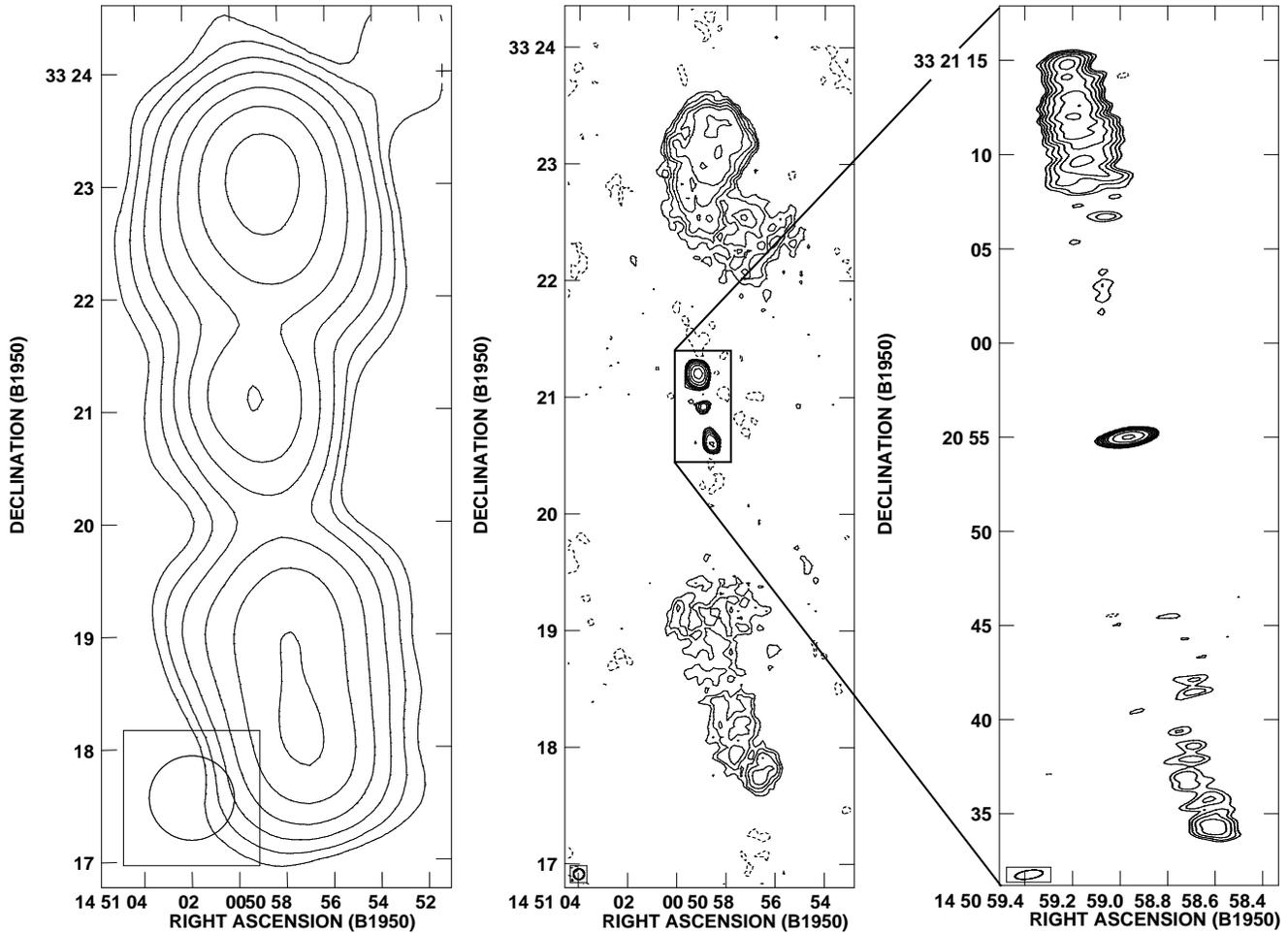,width=\textwidth,angle=270}
\caption{\label{fig:1450}Radio contour plots of the source
B\,1450+333. The FWHM size of the restoring beam is indicated in the
panel at the lower left side of each plot. The FWHM size of the
restoring beam is indicated in the panel at the lower left side of
each plot.{\bf a} Plot from the 1.4-GHz NVSS
survey. Contours are at ($-1.3$,1.3,2.6,5.2,10.4,20.8,41.6,93.2) mJy
beam$^{-1}$. {\bf b} Plot from the 1.4-GHz FIRST survey. Contours are
at ($-0.45$,0.45,0.64,0.9,1.27,1.8,3.6,7.2,14.4) mJy beam$^{-1}$. {\bf
c} Plot of the inner structure from our 5-GHz VLA
observations. Contours are at
($-0.12$,0.12,0.17,0.24,0.34,0.48,0.68,0.96,1.92,3.84) mJy
beam$^{-1}$.}
\end{figure*}

\begin{figure*}
\psfig{figure=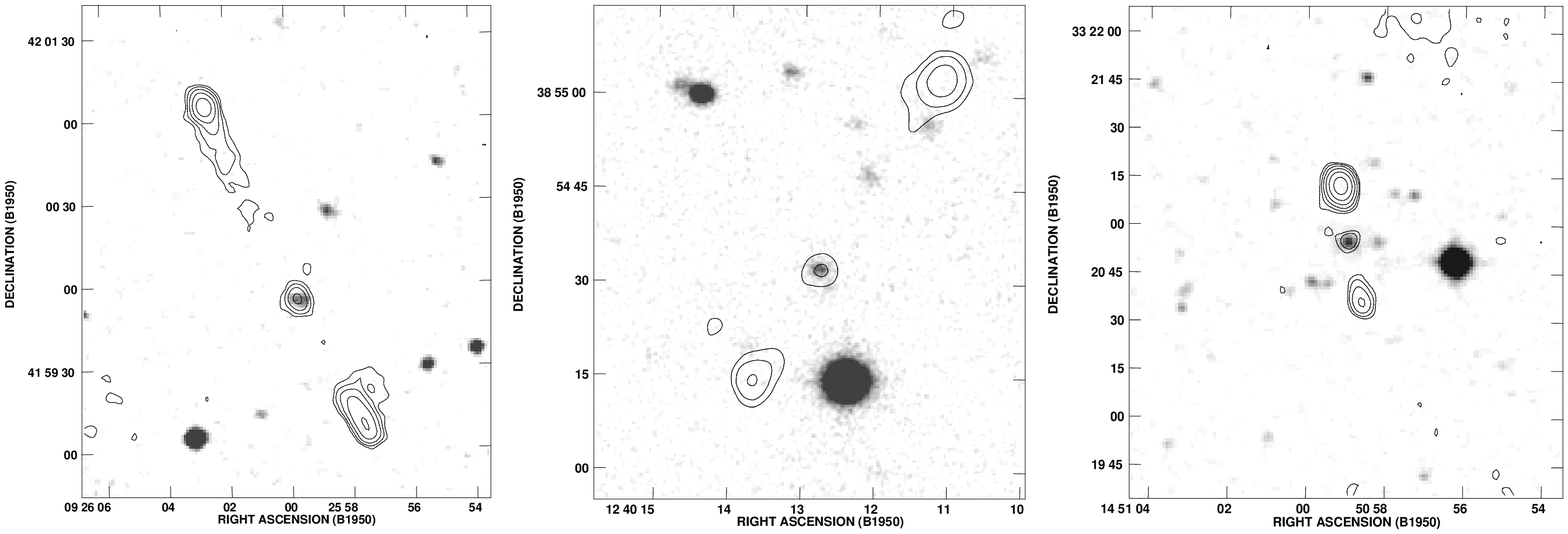,width=\textwidth}
\caption{\label{fig:all_opt}Overlays of the radio contours of the 1.4-GHz
FIRST survey and, in greyscale, optical images.
{\bf a} B\,0925+420. The optical image is from the POSSII-survey. 
{\bf b} B\,1240+389. The optical image is from a 1200s $R$-band observation
with the 1-m JKT telescope on La Palma. 
{\bf c} B\,1450+333. The optical image is from the POSSII-survey.}
\end{figure*}     

\section{The radio sources}
\label{sec:sources}
In this section we present radio and optical data of four sources we
have found in the 325-MHz WENSS survey (Rengelink et al. 1997) during
a search for large, extended radio sources. All four sources have
similar, peculiar radio morphologies.

\subsection{B\,0925+420}
\label{sec:0925}

B\,0925+420 (Fig. \ref{fig:0925}) appears as four well aligned radio
components in the 325-MHz WENSS and 1.4-GHz NVSS (Condon et al. 1998)
surveys. The higher resolution 1.4-GHz FIRST survey (Becker, White \&
Helfand 1994) has largely resolved out the outer two components,
indicating a relatively low surface brightness. The inner structure
has been well resolved by the FIRST survey, and shows a compact radio
core and two lobes with an edge-brightened (FRII-type; Fanaroff \&
Riley 1974) morphology.  The southern inner lobe is more compact than
the northern one, and is closer to the radio core. The radio core
coincides with a $R\approx18.3$ mag galaxy on the POSSII survey-plates
(see Fig. \ref{fig:all_opt}a); the magnitude has been taken from the
APM catalogue of the POSSI-survey and may be in error by 0.5 mag. 

We have taken an optical spectrum of this galaxy with the 2.5-m INT
telescope on La Palma on 8 April 1996. We used the IDS spectrograph
equipped with the R300V grating and a 1k$\times$1k TEK-chip. We have
used a slitwidth of $3\arcsec$, which yields a resolution of
$\approx\!15$\AA~in the dispersion direction. The resolution in the
spatial direction is determined by the seeing and is
$\approx\!1\,\farcs5$.  The total integration time was 1200s, split
into two 600s-exposures to allow cosmic-ray removal.  The resulting
spectrum (see Fig. \ref{fig:all_spec}a) shows a strong 4000\AA-break
and the $[$O{\sc ii}$]3727$ emission line for which we measure a
redshift of $0.365 \pm 0.002$. The projected linear size of the inner
source is $\sim\!800$~kpc, that of the outer source $\sim\!2450$~kpc.

\subsection{B\,1240+389}
\label{sec:1240}
The peculiar morphology of this source is best recognized on an
overlay of the FIRST and NVSS radio maps (Fig. \ref{fig:1240}).  The
inner structure of this source is only marginally resolved in the
FIRST survey. Nevertheless, it shows characteristics in common with
B\,0925+420, such as the diffuse outer lobes and the bright
two-sided inner structure. The slightly bend outer structure is also
suggestive of a Wide-angle tailed (WAT) radio source (e.g. Miley 1980,
O'Donoghue et al. 1990). However, in most WAT-sources, the inner
bright spots are not as compact as we see in this source; their
brightness decreases more gradually with increasing distance from the
radio core. Still, sensitive higher resolution observations of the
inner structure would be valuable to determine their radio morphology
in more detail.  Using a 1200s $R$-band CCD image obtained on 4
Feb. 1997 with the 1-m JKT telescope on La Palma, we have identified
the radio core with a $R\!\sim\!20.1\pm0.1$ mag galaxy (see
Fig. \ref{fig:all_opt}b). We have taken an optical spectrum of this
galaxy on 23 Feb. 1998 with the INT-telescope We have used the IDS
spectrograph, equipped with 1k$\times$1k TEK-chip and the
R158V-grating. We have integrated for 1800s using a slitwidth of
$2\arcsec$, which results in a resolution of $\sim\!20$\AA.  We
observe a 4000\AA-break at an observed wavelength of 5200\AA, yielding
a redshift of $0.30 \pm 0.01$\,.  The spectrum (see
Fig. \ref{fig:all_spec}b) does not show any emission lines above the
noise.  At a redshift of $0.30$, the projected linear size of the
inner structure is $\sim\!320$ kpc, that of the outer structure
$\sim\!860$ kpc. We remark that this source is the weakest case of the
four sources presented here.

\subsection{B\,1450+333}
\label{sec:1450}

The radio morphology of B\,1450+333 is best observed on the radio
map from the VLA FIRST survey. A contour plot of this map is shown in
Fig. \ref{fig:1450}b. The outer structure consists of a fat northern
radio lobe with no discernible hotspot, and a narrower southern lobe
with a weak, possibly double, hotspot.  We have observed the inner
structure at 5 GHz using the VLA in its BnA configuration. A
20-min. observation has been performed on June 28th 1998, using two
50-MHz bands centered at 4835 and 4885 MHz. Flux density calibration
was done using 3C\,286 as primary flux density calibrator and using
the Baars et al. (1977) values for its flux density.  The data have
been edited, mapped and phase self-calibrated using the {\sc aips}
data-reduction software package. The resulting radio map is shown in
Fig. \ref{fig:1450}c.  The inner radio lobes show an edge-brightened
morphology. Similar to the outer lobes, the southern inner lobe is
more extended and weaker than the northern inner lobe.  The southern
inner lobe lies on the radio axis defined by the core and the southern
outer lobe, but this is not the case on the northern side of the
source; here, the northern inner lobe lies well away from the line
connecting the core to the peak in intensity in the northern outer lobe.

\begin{figure}
\psfig{figure=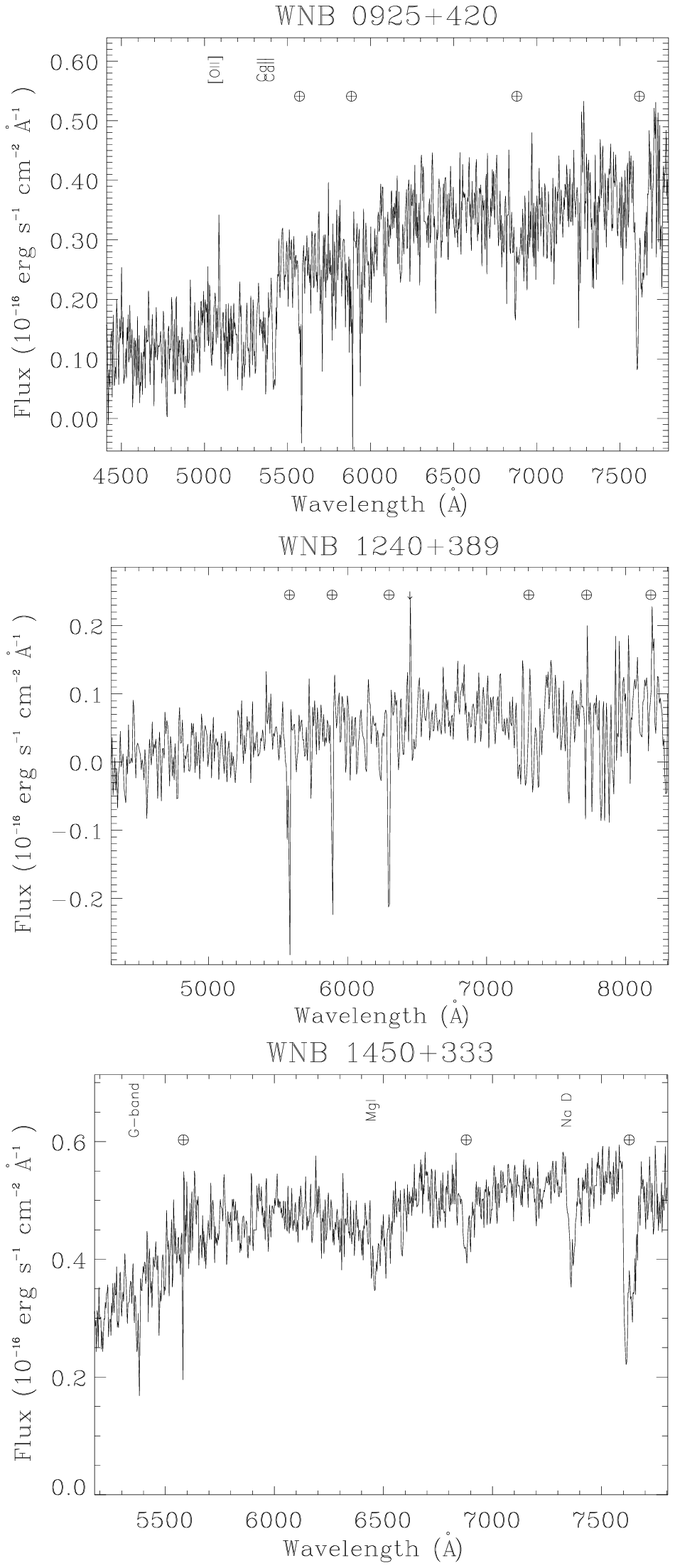,width=\columnwidth}
\caption{\label{fig:all_spec}Optical spectra of the host galaxies of B\,0925+420 (a),
B\,1240+389 (b) and B\,1450+333 (c). Identified lines have been
indicated on the plots. Atmospheric features have been indicated by a
`$\oplus$' symbol; those resulting from cosmic rays by a `$\downarrow$'
symbol. See the text for details.}
\end{figure}

We have identified the radio core with the brightest of an apparently
small group of galaxies on the POSSII-plates (see
Fig. \ref{fig:all_opt}c).  Its R-band magnitude, according to the APM
catalogue, is $\sim\!18.3$, with an estimated uncertainty of 0.5
mag. A spectrum of the host galaxy has been obtained by P. Best with
the 4.2-m WHT telescope on La Palma, using the red arm of the ISIS
spectrograph with the R150V grating and a 1k$\times$1k TEK-chip,
only. No dichroic has been used in these observations. The integration
time is 600s, using a 2$\arcsec$-wide slit. The resulting spectrum
(Fig. \ref{fig:all_spec}c) reveals no strong emission lines in the
range between 5200 and 8100\AA. Using the stellar Mg-b, Na-D and
G-band absorption bands, we determine a redshift $z = 0.249 \pm
0.002$.  The projected linear size of the inner source is $\sim\!186$
kpc, that of the outer source $\sim\!1700$~kpc.

\subsection{B\,1834+620}
\label{sec:1834}
The radio source B\,1834+620 has been observed by us at a variety of
frequencies and resolutions. Radio contour plots, as well as an optical
image of the host galaxy and its spectrum can be found in Paper
III. Our 8.4-GHz map unambiguously shows the
FR-II type morphology of the outer lobes. Of the two bright inner
knots, only the southern one is slightly resolved.  The 1.4-GHz map of
the inner structure shows that the southern inner component clearly is
not a knot in a jet but an edge-brightened radio lobe (see also Lara et
al. 1999). Its morphology
resembles that of the outer southern lobe in the 8.4-GHz map. The
northern inner component is more compact but also shows the structure
of an edge-brightened radio lobe. Higher resolution 5-GHz VLA
observations, also presented in Paper III, show this in somewhat more
detail.  The radio core coincides with a weak optical galaxy ($R_s =
19.7 \pm 0.1$) at a redshift of $0.5194 \pm 0.0002$, which is the
brightest member of a compact group of three galaxies. The projected
linear size is 1660~kpc for the outer source and 428~kpc for the inner
source.

\section{`double-double' radio galaxies}

\subsection{Definition}
\label{sec:definition}

The sources presented in Sect. \ref{sec:sources} are clearly different
from `standard' FRII-type radio galaxies. Since they consist of an
inner double-lobed radio structure as well as a larger outer
double-lobed structure, we have called these sources `double-double'
radio galaxies (DDRGs). In all the sources presented in this paper the
inner and outer radio sources are well aligned, to within $10\degr$.
However, there may be double-double sources where this is
not the case. Possible examples are the so-called X-shaped radio
sources (e.g. Leahy \& Williams 1984). In order to incorporate such
sources as well, we propose the following relatively general
definition of a DDRG: {\it A `double-double' radio galaxy consists of
a pair of double radio sources with a common centre. Furthermore, the
two lobes of the inner radio source must have a clearly extended,
edge-brightened radio morphology}.

The sources presented in this paper conform to this definition and are
examples of what one may call `aligned' DDRGs, i.e. sources where
the radio axes of the inner and outer structures line up very well.
We stress that for the inner structures a clear distinction has to be
made between `knots in a jet' and genuine radio lobes.  The definition
given above should discriminate against the former cases. For this,
high-resolution radio observations will usually be required.  

\begin{figure}
\psfig{figure=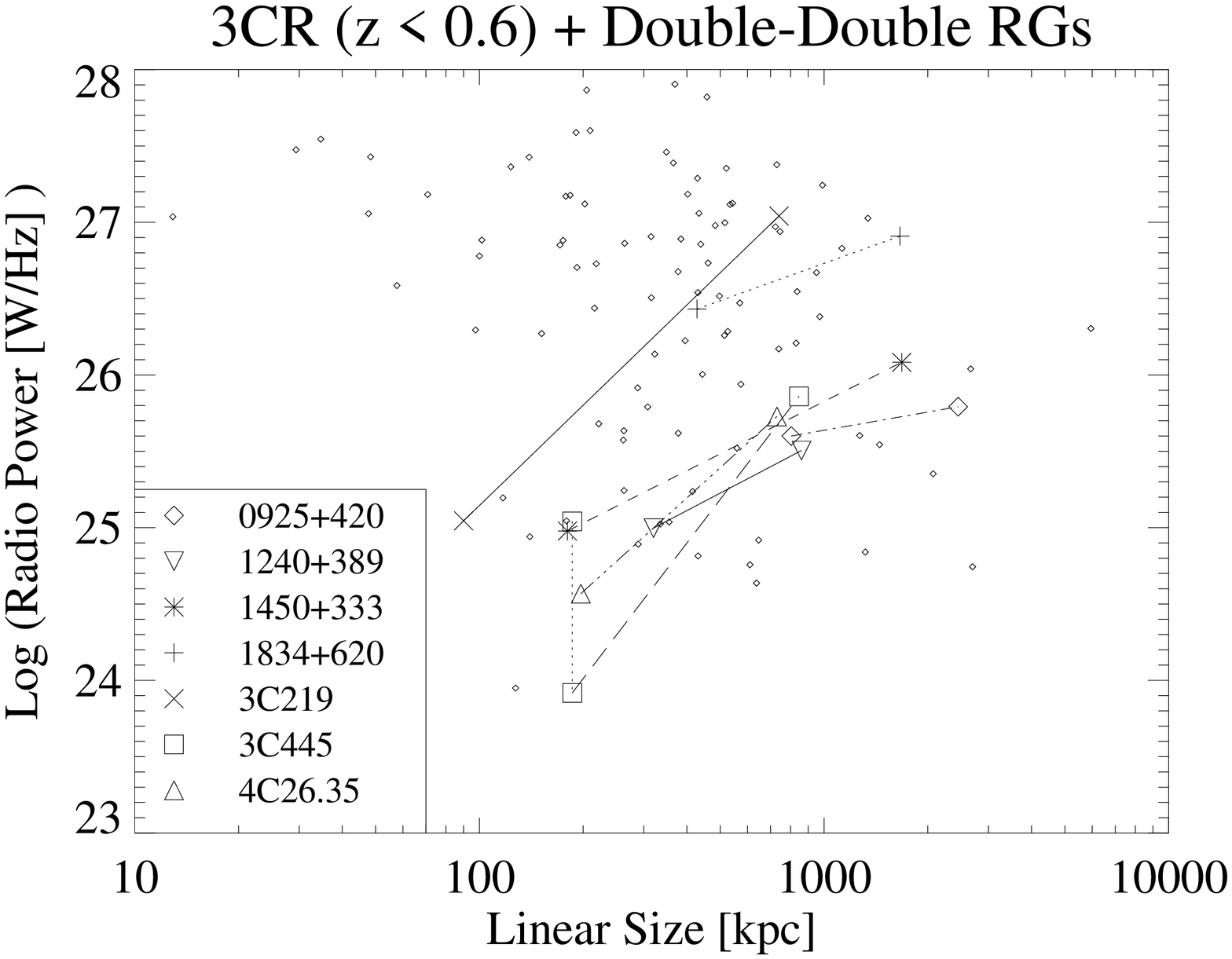,width=\columnwidth,angle=90}
\caption{\label{fig:p-d-diagram}Radio power - linear size ($P-D$) plot
of the DDRGs. The radio power is in W Hz$^{-1}$ at 1.4 GHz. For
comparison, we have also plotted $z<0.6$ 3CR sources in the LRL
subsample.  The larger symbols are the inner and the outer lobes of
the DDRGs. Points belonging to a single source have been
connected. For 3C\,445 we have plotted the lower and upper limit for
the flux density of the inner lobes, as discussed in the text.}
\end{figure}  

\begin{figure}
\psfig{figure=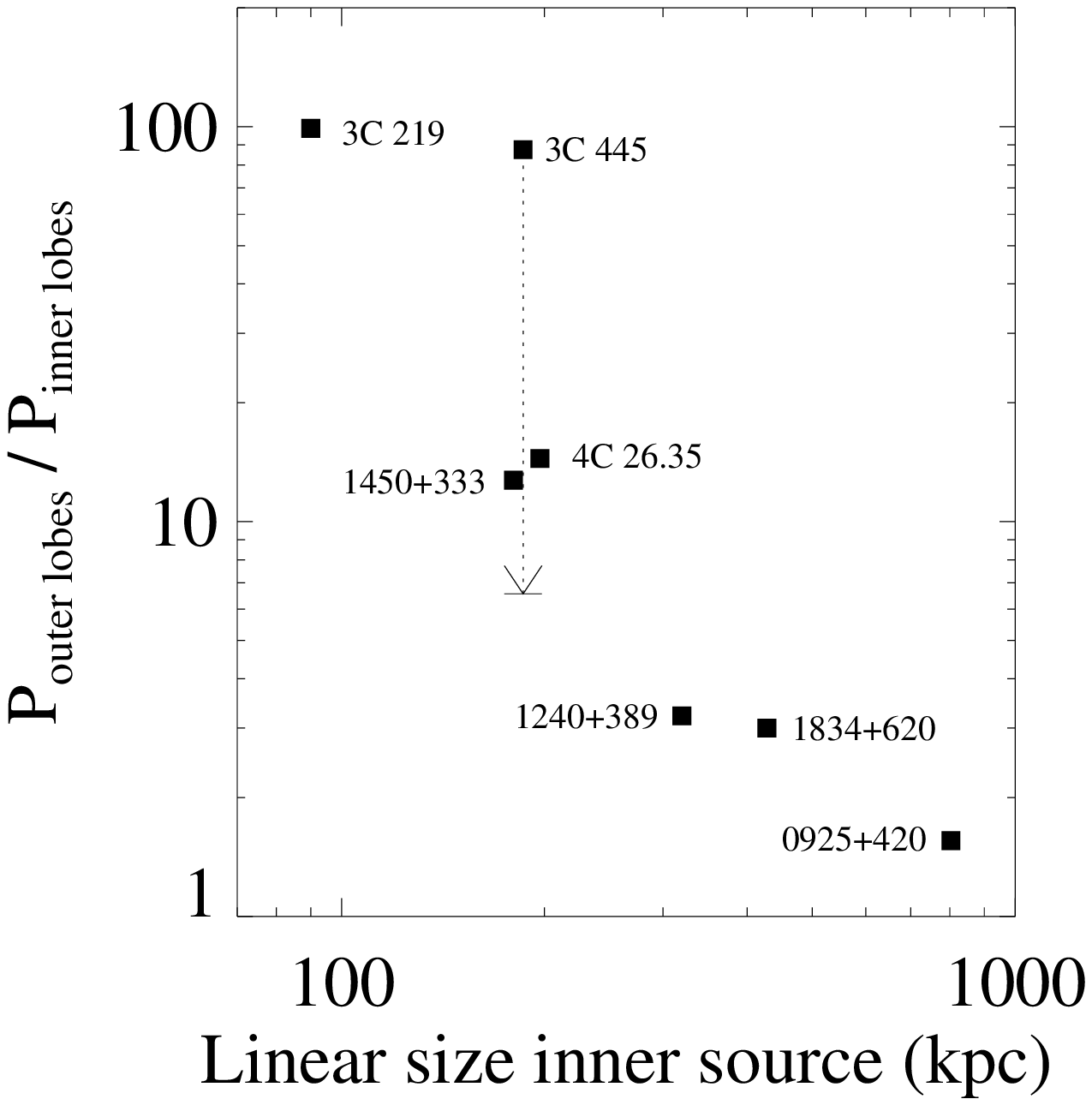,width=\columnwidth,angle=0}
\caption{\label{fig:p-d-plot}The ratio of the radio powers at 1.4~GHz
of the outer to the inner structures of the DDRGs, plotted against the
projected linear size of the inner structures. As mentioned in the
text, the flux density of the inner structure of 3C\,445 is not well
determined and we have plotted the point related to the lower flux
density limit only. The dotted arrow marks the range of possible
values for this source.}
\end{figure}

\subsection{Other DDRGs in the literature}
\label{sec:other_ddrgs}

Adopting the definition of DDRGs in Sect. \ref{sec:definition}, we
have searched the literature for more examples of aligned DDRGs and we
have found three more such sources. They are 3C\,445 (B2221-023;
Kronberg, Wielebinski \& Graham 1986, Leahy et al. 1997, Wil van
Breugel, priv. comm.), 4C\,26.35 (B1155+266; Owen \& Ledlow 1997)
and 3C\,219 (B0917+458; Clarke et al. 1992; Perley et al. 1994). All
three sources have well aligned inner and outer double radio lobes.
The source 3C\,219 has a highly asymmetrical inner
structure, both in armlength and in flux density. This may partly be
the result of an orientation effect. Its optical spectrum
(e.g. Lawrence et al. 1996) shows  broad emission lines
and a non-thermal contribution to the continuum. This is indicative of
sources whose radio axis may be close (i.e. $\la 45\degr$) to the
line of sight (e.g. Barthel 1989).  Still, its radio morphology agrees
to our definition. Also the source 3C\,445 is known to have broad
emission lines in its optical spectrum (e.g. Antonucci 1984).
However, in this case the asymmetry in the radio morphology and flux
density is much smaller.

We further note that other aligned DDRG-candidates are the X-shaped
radio source 4C\,12.03 (Leahy \& Perley 1991), which has a
small double-lobed inner radio source aligned with one of the two
pairs of outer radio lobes. The FRII-type radio galaxies 3C\,16 (Leahy
\& Perley 1991), 3C\,424 (Black et al. 1992) and the giant radio
galaxy B\,1545-321 (Subrahmanyan, Saripalli \& Hunstead 1996) also
have well aligned inner source structures.  Black et al. (1992)
present high-resolution observations of the inner structure of
3C\,424, which show that this source has, at least on one side of the
nucleus, an edge-brightened inner structure. On the other side of the
source the situation is less clear.  In the other two cases high enough resolution observations of the inner structures, necessary to determine
whether they have an edge-brightened morphology, are not available.

For this paper, we have decided to restrict our sample to the seven
sources for which we are confident that they are truly DDRGs and for
which both low and high resolution 1.4-GHz radio data are available to
measure their radio properties. These are the four sources introduced
in Sect. 2 and the sources 3C\,219, 3C\,445 and 4C\,26.35.  We stress
that this is by no means a uniform sample of DDRGs.


\section{Properties of the inner and outer components}
\label{sec:radiopowers}

We have measured the flux densities, projected linear sizes and
misalignment angles of the inner and the outer structures of the seven
DDRGs at 1.4~GHz. These are presented in
Tab. \ref{tab:source-properties}.  
For the source 3C\,445, the FIRST radio map does not allow a reliable
flux density measurements of the inner structure, since it is embedded
in the large-scale outer radio lobes which strongly affect the image
quality in the area of the inner structure.  The value given in
Tab. \ref{tab:source-properties} is therefore a lower limit. As an
upper limit we have taken the flux density of the same area of the
source measured in the NVSS survey.  Note that because of the extended
(outer) lobe emission in this region, this upper limit is a very
conservative choice.

We find that all these DDRGs have linear
sizes $\ga 700$ kpc.  Some DDRG candidates are very large
as well (3C\,16 ($D=504$~kpc), 4C\,12.03 ($D=820$~kpc) and B\,1545$-$345
($D=1320$~kpc) although other possible DDRGs such as 3C\,424 and
3C\,288 have a linear size of only 100 kpc and 170 kpc, respectively.  
The reason that the seven aligned DDRGs discussed here 
have a large size may partly be a selection effect: The
four sources introduced in Sect. \ref{sec:sources} have been selected to be of
Mpc-size.  However, the fact that none of the known complete samples of
smaller sized radio sources contain a large number of DDRGs 
suggests that these are not common among radio sources 
smaller than $\sim\!1$~Mpc.  Therefore, the
fact that all seven DDRGs have a very large linear size is most likely not a
selection effect but probably characteristic of the sub-class of radio sources
they form.

We have calculated the
1.4-GHz rest-frame monochromatic radio luminosities (the radio
power). We have assumed that all components emit isotropical and  we
have assumed power-law spectra with a spectral 
index $\alpha$ of $-0.75$ for the K-correction.  In
Fig. \ref{fig:p-d-diagram} we have 
plotted a radio power -- linear size ($P-D$) diagram of the inner and
the outer structures of the DDRGs. For comparison, we have plotted all
$z<0.6$ 3CR sources in the sample defined by Laing, Riley \& Longair
(LRL, 1983).  We find that, at 1.4~GHz, the inner components of all the
DDRGs presented here are less luminous than the outer components.  We
note that this also appears to be the case for the candidate DDRGs
3C\,16, 4C\,12.03\, and B\,1545-345.  Further, we find that in almost
all DDRGs, except for the source B\,1834+620, the radio power of the
inner component lies near or below the value separating the
radio-luminous FRII-type sources from the radio-weaker FRI-type
sources ($\sim 10^{25}$~W\,Hz$^{-1}$ at 1.4 GHz; Owen \& White
1991). Such low radio powers conflict with the edge-brightened
morphology of the inner lobes in these sources.  

In Fig. \ref{fig:p-d-plot} we have
plotted the ratio of the 1.4-GHz radio power of the outer lobes to
that of the inner lobes against the projected linear size of the inner
structure.  We find that the luminosity contrast between the inner and
outer structures appears to decrease with increasing size of the inner
sources.

\begin{table*}
\begin{minipage}{\textwidth}
\caption{\label{tab:source-properties} Some radio properties of the
seven DDRGs.  Column 2 gives the redshift; columns 3 and 4 give the
size of the inner and outer structures; column 5 gives the difference
in the position angle of the radio axes of the inner and outer
structure; columns 6 to 9 give the 1.4-GHz flux densities, $S$, and
radio powers, $P$, after subtraction of the radio core; columns 10 and
11 give the equipartition pressures, p, in the outer and inner
lobes. In the case of 3C\,445 the flux density of the inner component
could not be well determined from the maps of the FIRST survey; the
value given is a lower limit and we have omitted the equipartition
pressure calculation for the inner structure. The equipartition
pressures are calculated using the method given in Miley (1980),
assuming a cylindrical morphology and a filling factor of unity. They
are the average of the two sides of the inner or outer structure.}
\begin{tabular}{l c c c c c c c c c c c c c c c c}
\hline \multicolumn{1}{c}{(1)} & & (2) & & (3) & (4) & & (5) & & (6) &
(7) & & (8) & (9) & & (10) & (11) \\ \multicolumn{1}{c}{Source} & &
$z$ & & $\rm{D_{outer}}$ & $\rm{D_{inner}}$ & & $\Delta\theta$ & &
$S_{\rm outer}\,^a$ & $S_{\rm inner}$ & & $P_{\rm outer}$ & $P_{\rm
inner}$ & & p$_{\rm outer}$ & p$_{\rm inner}$ \\ 

& & & & $[$kpc$]$ &
$[$kpc$]$ & & $[\degr\,]$ & & $[$mJy$]$ & $[$mJy$]$ & &
\multicolumn{2}{c}{$[10^{25}$\,W\,Hz$^{-1}]$}& &
\multicolumn{2}{c}{$[10^{-13}$\,dyn\,cm$^{-2}]$} \\ 
\hline 
B\,0925+420 & & 0.365\phu & & 2450 & 803 & & 3 & & \phd99\phu & 63.7$^b$ & & 5.8 & \phs3.7 & & \phs1.7 & \phs7.6 \\ 
B\,1240+389 & & 0.30\phs\phu & & \phs860 & 320 & & 7 & & 24.1 & \phs7.8$^b$ & & 1.0 & 0.32 & & \phs1.2 & 40.7\\ 
B\,1450+333 & & 0.249\phu & & 1680 & 180 & & 7 & & \phs426\phu & 33.5$^b$ & & \php13 & 0.98 & & \phs1.4 & 61.3 \\
B\,1834+620 & & 0.519\phu & & 1660 & 428 & & 2 & & \phs604\phu & \php200$^c$ & & \php84 & \phps26 & & 16.7 & 84.7 \\ 
3C\,219     & & 0.174$^d$ & & \phs740 & \phs90 & & 2 & & 8045\phu & \phs\php90$^b$ & & 110 & 1.11 & & \phs9.6 & \php349 \\ 
3C\,445     & & 0.056$^e$ & & \phs846 & 186 & & 2 & & 5260\phu & \php$\ga$55$^b$ & & 7.3 & 0.08 & & \phs2.6 & \\ 
4C\,26.35   & & 0.112$^f$ & & \phs730 & 197 & & 2 & & \phs962\phu & 66.5$^b$ & & 5.4 & 0.37 & & \phs1.4 & \phps38 \\
\hline 
\end{tabular} \ \\
Notes: $a$ - Flux density measured from NVSS radio map minus $S_{\rm
inner}$; $b$ - Flux density measured from FIRST radio map; $c$ - Flux density
measured from our VLA observations; $d$ - Schmidt (1965); $e$ - Hewitt \&
Burbidge (1991); $f$ - Owen, Ledlow \& Keel (1995)
\end{minipage}
\end{table*}

\section{Discussion}
\label{sec:discussion}

\subsection{Formation scenarios}

The two-sidedness and relatively high degree of symmetry of the inner
radio structures of the DDRGs strongly points towards a causal
connection of the development of the inner structure with processes in
the nucleus.  Any model for the formation of these sources should, at
least, be able to explain the following properties. First, for the
inner structure, it should explain:
\begin{enumerate}
\item The two-sidedness and relatively good armlength symmetry,
\item the alignment of the radio axis with the outer lobes,
\item the edge-brightened radio morphology, and
\item the relatively low luminosity as compared to the outer lobes.
\end{enumerate}
Further, in connection to the outer structure, it should clarify:
\begin{enumerate}
\setcounter{enumi}{4}
\item The large size of the outer structures, and
\item the possible presence of hotspots in the outer lobes, such as seen in B\,1834+620. 
\end{enumerate}
Here, we will discuss three scenarios that may explain the existence
of the inner lobes in our DDRGs, all of which have been used before to
explain or predict radio structures on different size scales.

\subsubsection{A change in the jet outflow direction}

A new radio source may be started by a sudden change in the outflow
direction of the jet. At first the redirected jet will traverse the
cocoon, through which it will travel almost ballistically (e.g. Clarke
et al. 1992, paper II), but eventually it will reach the cocoon
boundary and run into the much denser IGM again. From that point on a
new hotspot and lobe structure may be created. This may result in
the formation of an X-shaped radio source (e.g. Leahy \& Williams
1984; Parma, Ekers \& Fanti 1985).  
For the DDRGs presented here,
however, we consider such a redirection of the jet flow unlikely to
be the cause of the observed source structures, because of the good
alignment of the inner and the outer sources.

\subsubsection{Backflow instabilities}

Structures whose radio morphology may resemble the DDRGs have been
found in the numerical simulations of Hooda, Mangalam \& Wiita
(1994). They simulated the evolution of extra-galactic radio jets on
time scales corresponding to $10^8$ yrs and which propagate out to
$\sim\!400$~kpc. They find that around the time at which the
propagation of the head of the jet becomes nearly subsonic,
instabilities arise in the backflow of the radio lobe. These
instabilities eventually pinch off the jet channel and thus disconnect
the outer lobe structure from the jet flow.  As a result the original
hotspot, and eventually the whole outer lobe, will fade. On the other
hand, the pinching of the jet channel gives rise to the development of
a new shock front, which after a while may form a new radio
lobe-like structure.  On the basis of the results of these simulations
Hooda et al. predict a population of large radio sources with diffuse outer
lobes and edge-brightened structures interior to these.

Since the jet is pinched off, a good alignment between inner and outer
radio structures is a natural consequence.  However, in order to pinch
the jet channel, instabilities are required which must be strong
enough to achieve this. It is rather surprising that something to that
effect would happen on both sides of the radio core at roughly the
same distance and also at approximately the same time (within a few
percent of the source age). Especially since the outer structure of a
source like B\,1450+333 is highly asymmetric in morphology, so that any
backflow in these lobes will have very different properties in each
lobe.

Therefore, although this scenario may be able to explain the
formation of a one-sided inner structure, we find it unlikely that it
can explain the two-sided and symmetric nature of the inner structures
of the DDRGs.  Note that in the case of 3C\,219, which has a highly
asymmetrical inner structure, this objection carries less weight.

\subsubsection{Interrupted jet activity}

Short term variations in the energy output are known to occur in
almost all AGN. In radio-loud AGN, small changes in the jet power
may lead to shocks which are visible as discrete `blobs' in the jet
(e.g. Rees 1978), or become manifest only once they reach the hotspot
and change its luminosity.  For example, the much larger asymmetry
seen in hotspot luminosities, as compared to lobe luminosities, in
FRII-type radio sources (e.g. Macklin 1981, Paper III)
suggests that jet powers vary significantly during the
lifetime of a radio source (see paper III).  It is plausible that such
variations occur on a large range of time scales, without seriously disrupting
the jet flow.  

A complete halting of the acceleration of the jet production in the
central engine will be desastrous for the jet channel: The loss of
pressure, assumed to be provided by the jet material flowing through
the channel (e.g. Kaiser \& Alexander 1997), will result in its
collapse.  If the nuclear outflow restarts at some time after the
`old' jet channel has disappeared, it will have to clear out (or
`drill') a new channel through the cocoon. This may result in the
formation of a shock at the head of the jet which may manifest
itself as a hotspot, under the restriction that the density of the
cocoon is high enough to allow such a shock to form.

Analytical models (e.g. Cioffi \& Blondin 1992) and numerical
simulations (e.g. Clarke \& Burns 1991, Loken et al. 1992) predict
that the density inside a cocoon is much lower (up to a factor of 100)
than that in the original unshocked InterGalactic Medium (IGM), which
is too low for the formation of a strong jet shock.  However, in Paper II
we present a model that allows the density in the old cocoon to be
much larger than previously assumed, and as a result allows the
formation of inner hotspots after the jet has restarted.  We show that
such a model can furfill the first five requirements mentioned at the
beginning of this section. Whether the requirement of the allowed
presence of hotspots in the outer lobes is met depends on the length of
time during which the jet production is interrupted; we will return to
this topic later.

Only the source 3C\,219 cannot be easily explained in this
scenario, since the two oppositely directed jets are assumed to
restart at more or less the same time whereas the inner structure of
3C\,219 is highly asymmetric in armlength; as mentioned above, another
mechanism may be at work in this source. For the other sources
presented here, however, interruption of the jet formation is the most
promising scenario for their formation.

\subsection{Causes and consequences of the interruption}

\subsubsection{Time scales of the interruption of the jet activity}
\label{sec:timescales}

A constraint on the time scale between the halting and restarting of
the jet can help in determining the cause of the interruption.  In all
DDRGs the outer lobes have not yet faded away; unfortunately, this
does not provide a very tight constraint, since it is not clear at all
for how long a radio lobe will remain detectable after the jet has
stopped supplying it with energy.  Kommissarov \& Gubanov (1994)
estimate fade-away time scales of several $10^7$ yr for a small sample
of currently inactive (or `relic') radio sources, which is comparable
to the time scale of the activity itself. Therefore, this estimate
provides only an upper limit to the time scale of the interruption.

The best example of a radio source in which the interruption must have
been relatively brief is B\,1834+620, since it still shows a bright and
compact hotspot in one of its lobes (Paper III). This indicates that
remnants of the `old' jet must still be arriving in the hotspot, or at
least until very recently; the fade away time of hotspots of size $\sim
10$ kpc is estimated to be a few $10^4 - 10^5$ yr (e.g. Clarke et
al. 1992, Paper II).  Assuming that the radio axis of this radio galaxy
has an orientation angle of $\le 45\degr$ to the plane of the sky
(cf. Barthel 1989) limits the time elapsed since the interruption to
less than between 1.1 and 6.4 Myr (depending on the orientation; see
Paper III).  Since in this amount of time also the inner structure must
have formed and grown to its current size,  the time scale of the
interruption of the jet must have been close to 1 Myr, at
most. Although B\,1834+620 is only one example, it shows that if there
were a common mechanism for the interuption of the jet activity, it
limits the time scale of the interruption to a few Myr, only.
Further, it appears as if it only starts to operate after an elapsed
activity time of a few times $10^7$ yr, since these are the estimated
ages of sources as large as the outer structures of DDRGs (e.g. Mack et
al. 1998, Schoenmakers 1999).  However, the model we present in Paper
II suggests that this later requirement is a selection effect: If the
interruption occurs much earlier in the lifetime of a radio source, the
restarted jet may not be able to form the inner lobes due to a still
too low density of the cocoon. Therefore, a DDRG would only form when
the outer source has grown to a large size.  However, if this were the
case and the time elapsed between the start and the interruption of a
jet is much smaller than the total lifetime of the radio source, then
multiple interruptions during a sources lifetime may occur and it can
be expected that DDRGs would be more common among large radio sources.

The DDRGs are not the only radio sources that show evidence for interrupted activity on
relatively short time scales.  Several compact and therefore presumably
young radio sources have been found that are clearly associated with
larger scale radio emission (e.g. B\,0108+388, Baum et al. 1990,
Owsianik, Conway \& Polatidis 1998; B\,1144+352, Schoenmakers et
al. 1999a; B\,1245+676, de Bruyn et al. in preparation). The radio
structure of these sources can be interpreted as the result of an
interruption of the radio activity.  If the cause of the interruption not only
affects the radio jet formation, but the whole central activity, then
it is reasonable to assume that radio-quiet AGN show similar
interruptions.  Although these sources constitute $\sim\!90\%$ of the AGN
population, a direct observation of the interruption in these objects
is very difficult since there is no good tracer of past activity on
timescales of a few Myr.

\subsubsection{Scenarios for the interruption of the jet activity}

In AGN, the formation and properties of jets must be closely related to
the properties of the central black hole and its accretion disk,
although it is still largely unclear how this is achieved.  Assuming
that the jet formation is related to the accretion flow, an
interruption of the jet formation for a period of a few Myr is most
likely caused by a passing event which temporarily disturbs the
stability of the accretion disk.  There a several scenarios to achieve
this, among which the following:

First, an internal instability in the accretion disk, for instance due
to radiation pressure induced warping (Pringle 1997, Natarajan \&
Pringle 1998).  This does not require an external cause for the
interruption, but it is unlikely to provide the mechanism needed to
explain the DDRGs. First, it is largely unclear what the timescales are
for such a process (Pringle 1997). Also, the occurence of a warping
instability is likely to change the direction of the jet considerably
(Natarajan \& Pringle 1998). Natarajan \& Pringle state that to produce
a directional stable jet the central black hole must be spinning, but
this also suppresses the formation of a warping instability needed to
interrupt the jet formation. Therefore, this mechanism is unlikely to
operate in the DDRGs.

Second, a large cloud of gas may fall into the centre of the
galaxy. This may also disturb the stability of the accretion disk and
thus the jet formation. However, it is hard to envision a relation
between possible arrival times of such a cloud in the AGN and the large
size of the DDRGs unless the arrival of the gas is somehow
regulated. The mechanism causing the halting of the jet operates on a
typical time scale of a few $10^7$ yr after the jet activity first
started. This suggests the following mechanism.  If we assume that AGN
activity is triggered by an interaction or merger, numerical 
simulations of colliding galaxies show that these usually do
not merge completely in the first encounter (e.g. Barnes \& Hernquist
1996). Although the infalling galaxy loses a large fraction of its gas
and stars to the main galaxy, parts of it pass the main
galaxy. These parts will turn around and collide again with the main
galaxy. Simulations show that a typical merger is only complete after two or
more of such encounters, which may take up to a few $10^8$ yr. Each 
passage through the host galaxy might lead to a new phase of increased
kinematical instabilities in the host galaxy. 
The turn-around time scale of the infalling galaxy is roughly comparable
to a rotation time of the main galaxy (i.e. a few $\sim10^7$ yr; Barnes
\& Hernquist 1996) which agrees with the estimated time scale between
the first start and the ceasing of the jet activity in the DDRGs.
On the other hand, substructures in the infalling gas flow may also cause
instabilities in the accretion flow; these would then be related to
phenomena operating on much smaller timescales than the turn-around time scale.

Although the notion that AGN activity may be triggered by a merger
event has been around for a long time, there is no solid proof for
this, yet. 
It is not clear at all how and in which state the gas
flows into the inner few tens of pc around the AGN. The scenario we
propose depends on a phenomenon which must be quite common in galaxy
mergers and interactions, namely multiple encounters between the two
galaxies involved. If the first encounter triggers the AGN and
subsequent encounters destabilize it, then interruption of the activity
must be common among AGN. This may lead to many more sources that are
in a restarted phase than are actually observed. The exact behaviour of
two merging galaxies depends strongly on parameters as their mass
ratio, rotation, relative velocities and orbits.  Discussing this in
detail extends beyond the scope of this paper, though.

\subsection{Possible consequences for the formation of giant radio sources}

Some radio sources can grow to sizes of a few Mpc (e.g. Saripalli et
al. 1986, Subrahmanyan et al. 1996, Mack et al. 1998, this paper). It
has often been suggested that these Giant Radio Galaxies (GRGs) must be
very old and/or in extremely low density environments. But most
indications for this have been obtained by measuring spectral ages
(e.g. Mack et al. 1998, Schoenmakers et al. 1999d), which may be
inaccurate (e.g. Eilek 1996), and by depolarization studies, which are
often difficult to interpret (e.g. Schoenmakers et al. 1998).

An alternative formation scenario for these large sources,  proposed
by, e.g., Subrahmanyan et al. (1996), is that they result from multiple
periods of jet activity. The DDRGs presented here suggest that such a
formation scenario for GRGs may indeed be valid: If the inner lobes
continue to advance at the expected high velocities (see paper II and
III), they will rapidly reach the boundary of the outer lobes and the
radio source will evolve further.

In Sect. \ref{sec:timescales} we have discussed several Mpc-sized
radio sources which also show inner lobes on much smaller scales
(i.e. a few tens of pc).
Sources such as these
may have restarted their jet production only recently and may therefore
be progenitors of the DDRGs presented in this paper.  Another possible
example is the currently largest radio source known in the Universe,
3C\,236, with a projected linear size of 5.7 Mpc (e.g. Strom \& Willis
1980). It has a bright radio core showing a complex, multiple and
apparently young structure on pc-scale (e.g. Barthel et al. 1985); it
is not unlikely that this core structure is the result of interrupted
jet activity, as well.

The large number of Mpc-sized radio sources without signs of
interruption shows that this phenomenon is probably not necessary to
obtain such a large size.  Also, a single prolonged period of activity
would probably be more efficient in producing a large radio source than
one with repeated interruptions of the jet formation.  Only in the case
that the AGN has run out of fuel, and that a restart results from a new
supply of energy, a larger radio
source may emerge than that formed during the first phase of activity. The
requirement that the outer lobes must still be visible forces such a
restart to occur within a few $10^7$ yr, which may be achieved by the
same mechanism we propose for the DDRGs. Whether this would work
repeatedly is hard to imagine, though. Improved statistics on interrupted
activity, both in GRGs and in smaller sized sources, may help us to better
understand the influence and importance of this process in radio source
formation.

\section{Conclusions}

We have discovered four Mpc-sized radio sources which consist of two
aligned, but unequally sized, FRII-type radio sources with a coinciding
radio core. We have called these sources `double-double' radio galaxies
(DDRGs) and have presented a general definition for this class of radio
source.  In the literature we have found three more sources (3C\,219,
3C\,445 and 4C\,26.35) that conform to our definition and for which
1.4-GHz radio data is available to us. We have included them into a
small sample of seven DDRGs with well aligned inner and outer radio
structures.

We find that all these DDRGs are large radio sources, with linear sizes
exceeding 700 kpc. Further, we find that in all cases the inner radio
structure is less luminous at 1.4 GHz than the outer structure and the
luminosity contrast appears to decrease with increasing linear size of
the inner structure.

We have discussed several scenarios for the formation of the inner
structures and we conclude that an interruption of the jet production
in the AGN is the most likely scenario.  The detectability of the outer
radio structures and the presence of a hotspot in one of the outer
lobes of the DDRG B\,1834+620 indicate that the time scale of the
interruption must be small, a few Myr at most.  Further, we have
discussed scenarios for the cause of the interruption.  We conclude
that multiple encounters between interacting galaxies is a likely
scenario, but we note that nothing is known in detail about the
consequences of such encounters for AGN activity.

Notwithstanding the cause of the interruption, we believe that DDRGs
provide excellent evidence for small time scale (i.e. a few Myr)
interruptions of the jet activity in AGN.  Detailed studies of the
DDRGs are of key importance to learn more about this phenomenon, about
duty cycles of AGN and how this affects the evolution of a double-lobed
radio source.

\section*{Acknowledgments}

The authors would like to thank P. Best and M. Lehnert for their input
and many helpful discussions. P. Best is also thanked for obtaining the
spectrum of the host galaxy of B\,1450+333.  Further, we thank the
referee, J.P. Leahy, for helpful comments that
improved this work considerably.  The INT and WHT are operated on the
island of La Palma by the Isaac Newton Group in the Spanish
Observatorio del Roque de los Muchachos of the Instituto de Astrofisica
de Canarias. The Westerbork Synthesis Radio Telescope (WSRT) is
operated by the Netherlands Foundation for Research in Astronomy (NFRA)
with financial support of the Netherlands Organization for Scientific
Research (NWO).  The National Radio Astronomy Observatory (NRAO) is
operated by Associated Universities, Inc., and is a facility of the
National Science Foundation (NSF).  This research has made use of the
NASA/IPAC Extragalactic Database (NED) which is operated by the Jet
Propulsion Laboratory, California Institute of Technology, under
contract with the National Aeronautics and Space Administration.  The
Digitized Sky Surveys were produced at the Space Telescope Science
Institute under U.S. Government grant NAG W-2166. The images of these
surveys are based on photographic data obtained using the Oschin
Schmidt Telescope on Palomar Mountain and the UK Schmidt Telescope. The
plates were processed into the present compressed digital form with the
permission of these institutions.  This work was supported in part by
the Formation \& Evolution of Galaxies network set up by the European
Commission under contract ERB-FMRX-CT96-086 of its TMR programme.

{}

\end{document}